\documentclass[aip,jmp,article,reprint,onecolumn,numerical]{revtex4-1}
\usepackage{graphicx}
\usepackage{latexsym}

\usepackage{amsmath}
\usepackage{amssymb}

\usepackage{amsthm}

\begin{document}

\title{New Class of 4-Dim Kochen-Specker Sets}

\author{Mladen Pavi\v ci\'c\footnote{E-mail: pavicic@grad.hr}}

\affiliation{Institute for Theoretical Atomic, Molecular,
and Optical Physics at Physics Department at
Harvard University and Harvard-Smithsonian Center for
Astrophysics, Cambridge, MA 02138 USA \&\ Chair of Physics, Faculty of Civil Engineering,
University of Zagreb, Zagreb, Croatia.}

\author{Norman D.~Megill\footnote{E-mail: nm@alum.mit.edu}}

\affiliation{Boston Information Group, 19 Locke Ln., Lexington,
MA 02420, U.~S.~A.}

\author{P.\ K.\ Aravind\footnote{E-mail: paravind@wpi.edu}}

\author{Mordecai Waegell\footnote{caiw@wpi.edu\\  \\
{\Large{{\it Journal of Mathematical Physics}, {\bf 52}, No.~2 (2011).}}}}

\affiliation{Physics Department, Worcester Polytechnic Institute,
Worcester, MA 01609, U.~S.~A.}


\begin{abstract}
We find a new highly symmetrical and very numerous class (millions of
non-isomorphic sets) of 4-dim Kochen-Specker (KS)
vector sets. Due to the nature of their geometrical symmetries,
they cannot be obtained from previously known ones.
We generate the sets from a single set of 60
orthogonal spin vectors and 75 of their tetrads
(which we obtained from the 600-cell) by means of
our newly developed  {\em stripping technique}.
We also consider {\em critical KS subsets} and analyze
their geometry. The algorithms and programs for the generation
of our KS sets are presented.
\end{abstract}

\smallskip


\maketitle

\section{\label{sec:intro}Introduction}

Kochen-Specker (KS) sets have recently been studied extensively
due to new theoretical results which have prompted new
experimental and computational techniques.
The theoretical results consider conditions under
which such experiments are feasible
\cite{cabell-02,barrett}, including  single qubit KS setups
\cite{cabell-03,cabello-moreno-07}. Such results and
experiments are applicable to quantum computational
rules of dealing with qubits and qutrits within a large
number of quantum gates.

The experiments were carried out for
spin$-\frac{1}{2}\otimes\frac{1}{2}$ particles (correlated
photons, or neutrons with spatial and spin degrees of freedom), and
therefore in this paper we provide results only for 4-dim
KS vector sets of yes-no questions (KS sets for short).
Recent designs \cite{cabello-08} and experiments
\cite{cabello-fillip-rauch-08,b-rauch-09,k-cabello-blatt-09,amselem-cabello-09,liu-09,moussa-laflamme-10}
deal with state-independent vectors. Such
experiments can tell us a great deal more about
quantum formalism and the geometry involved in
obtaining new functional KS setups and quantum
gate setups in general. On the other hand, a recent result
of A.~Cabello \cite{cabello-10} connects noncontextuality
and therefore the KS theorem with quantum nonlocality
and opens the possibility of using KS sets in
quantum information experiments.

For both of these applications, it is
important to have many nonisomorphic critical (empirically
distinguishable) KS sets. Before our work, only eight 4-dim ones were
known (seee below). In this work, we present thousands of new
nonisomorphic 4-dim critical and millions of non-critical
KS sets. This is also important for a better understanding
of quantum systems. First, it seemed that quantum
gates and system state configurations that would allow
only quantum representations were very sparse.
We show that they are actually abundant.
Second, the configurations of Hilbert space vectors and
subspaces in KS sets have interesting symmetries that have
intrigued many authors since the discovery of the KS theorem.~\cite{zimba-penrose,penrose-02,peres-book,massad-arravind99,aravind-ajp,mermin93,topos2,conway-kochen-02,ruuge05,brunet07,blanchfield-10}
We now get many new symmetries
because we obtain a new disjoint class of KS sets, none of
which were previously known and none of which can
be built up from previously known ones.

Recently, we found that all known KS sets with up to 24 vectors and
component values from the set \{-1,0,1\} (including one with 18 vectors
for which experiments have been carried out) can be obtained from a
single KS set with 24 vectors and 24 tetrads (blocks) originally found
by A.~Peres.~\cite{pmm-2-09}  The following is a brief summary of the
techniques we used for that discovery.

When a KS set can be obtained from a larger one by stripping (removing)
tetrads, the stripped tetrads correspond to redundant detections within
a measurement of spin projections.  ``Critical'' sets are the smallest
empirically distinguishable KS sets in the sense that they cannot be
reduced to each other by stipping tetrads.  Instead, stripping any tetrad
from a critical set will cause the set to cease to be a KS set.

Using the ``stripping technique'' (which
we explain below), we exhaustively generated all
possible 24-24 McKay-Megill-Pavicic (MMP) 
hypergraphs\cite{bdm-ndm-mp-fresl-jmp-10}
[each vertex in a hypergraph represents a vector (state) in a Hilbert
space and each tetrad (block) corresponds to four mutually orthogonal vectors]
and found (after several months of computation on our
500 CPU cluster {\em Isabela}) that among over $10^{10}$
nonlinear equations to which the hypergraphs correspond, only
one has a solution.~\cite{mp-report-10}
The solution is, of course, isomorphic to Peres'
24-24 KS set mentioned above. Therefore we named this set
simply the ``24-24 KS set'' and the family of KS sets that can be
obtained from it, the ``24-24 KS class.''

We obtained KS subsets of the 24-24 set by a {\em stripping
technique}, which consisted of stripping blocks off of
the initial 24-24 KS set then checking whether the resulting
subset continued to be a KS set. There are altogether 1232 such KS
subsets.~\cite{pmm-2-09} Looking at these sets as sets of vectors
with component values from a \{-1,0,1\} that do not allow a numerical
evaluation (KS theorem!), we might not see any apparent reason why,
among trillions of instances, there would not also be KS
sets that are not subsets of the 24-24 set. But it turns out that
there is none. We prove that by an exhaustive generation of KS sets
with 18 to 23 vectors. They are isomorphic to the subsets of the
24-24 set.~\cite{mp-report-10}.

When we strip off blocks, we eventually reach  smallest KS sets
in the sense that any of them ceases
to be a KS set if we strip off an additional block.
We call such smallest sets {\em critical sets}. This
is the definition we used in Ref.~\onlinecite{pmm-2-09},
and it differs from the definition we used in
Ref.~\onlinecite{aravind10}, which is based on deleting
vectors (directions, rays). In Ref.~\onlinecite{pmm-2-09} we
proved the there are altogether six critical KS subsets in the
24-24 KS class. These are:
18-9,\cite{cabell-est-96a} 20-11,\cite{kern}
another 20-11,\cite{pmmm03a}, two
22-13s,\cite{pmmm03a} and 24-15.~\cite{pmm-2-09}

The main focus of this paper is to describe another isolated family of
KS sets (the 60-75 KS class), which we discovered by means of our stripping
technique applied to a KS set
with 60 vectors and 75 blocks.  This latter set was obtained
from the 4-dim polytope called the 600-cell.\ \cite{aravind-600}
We call this new family the ``60-75 KS class.'' It turns out that it contains
millions of KS sets and thousands of critical
sets, and this is what is novel in this paper.~\cite{mp-vienna-talk-10}
Previously, we found only several isolated examples from this
class using a different technique.~\cite{aravind-600,aravind10}
Being based on the geometry of the 600-cell, the 60-75 set
provides us with highly symmetrical configurations of
60 rays, and this symmetry can be traced down to its smallest
(with an estimated a confidence of over 95\%) 26-13 KS subset (described in
Sec.~\ref{sec:critical}).  In particular,
the smallest maximal loop of any KS set from the 60-75 class
forms an octagon, while all sets from the 24-24 class have
a hexagon maximal loop. The fact that the smallest KS set from the
60-75 class has 26 vectors,  together with the fact that we could not
build up to any subset of the 60-75 class with 24 vectors proves the
disjointness of the two classes with a confidence of over 95\%.

Below, we  present some of many (several thousand) new
critical sets  from the 60-75 KS class, give
the algorithms that
enabled us to find them, and investigate their symmetry
and geometry, comparing it with the ones of the 24-24 KS class.

\section{\label{sec:critical}Critical Sets}

A discovery we made for the 18 through 24 vector sets was
that the maximal loop of edges in all their MMP hypergraphs
was a hexagon and that in all of them there was only one hexagon.
(In an MMP hypergraph vertices correspond to vectors and  edges
to tetrads. So, {\tt 1} denotes the 1st vector, \dots , {\tt 9} the 9th,
{\tt A}, the 10th, \dots , {\tt Z} the 35th, {\tt a} the 36th,
\dots {\tt y} the 60th.) That gave us the idea of the stripping
technique, because stripping the edges in a way that preserves
the hexagon might give a comparatively small number of subsets
and critical sets. That was confirmed in  Ref.~\onlinecite{pmm-2-09}.

\begin{figure}[htp]
\begin{center}
\includegraphics[width=0.248\textwidth]{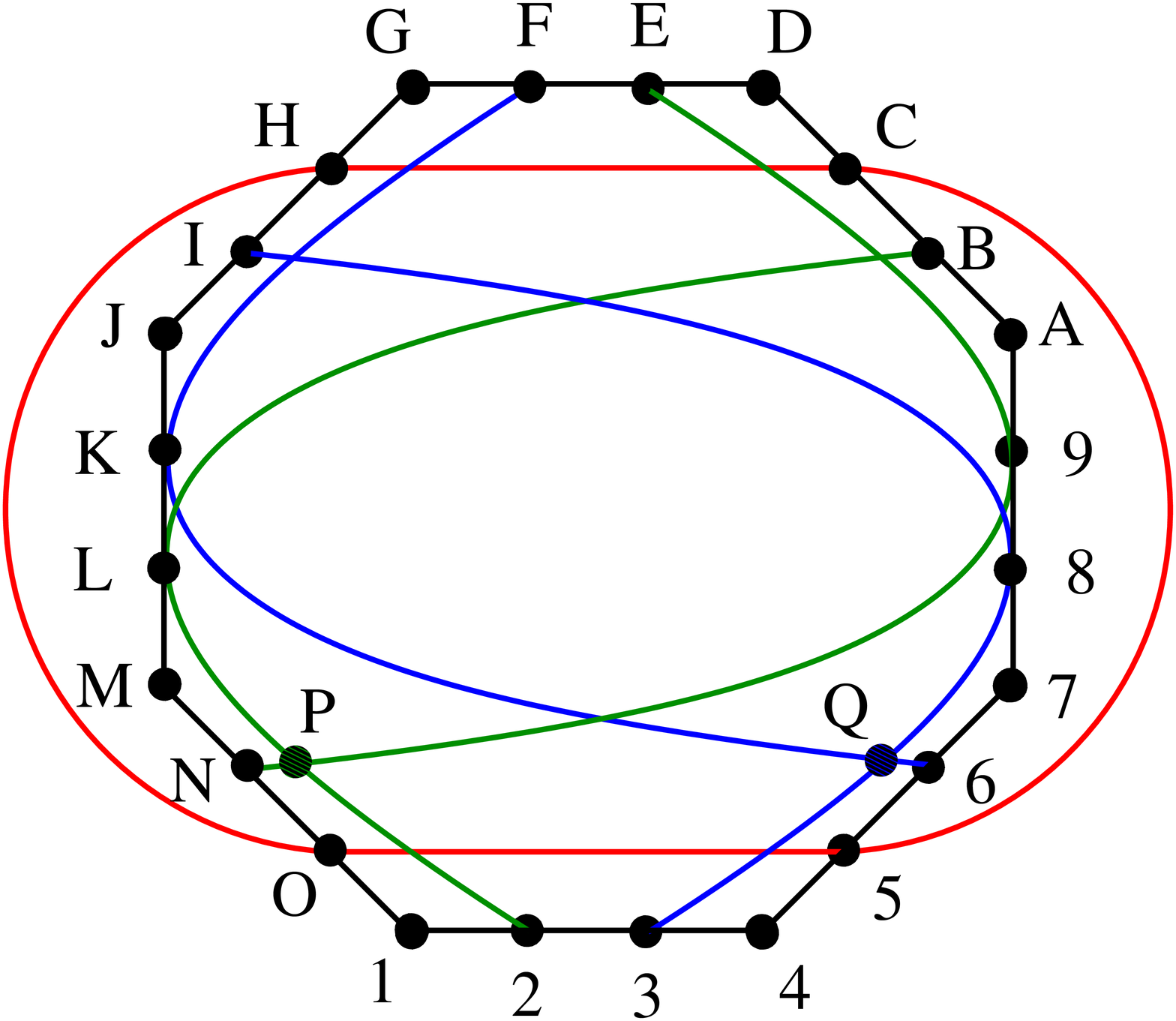}
\includegraphics[width=0.263\textwidth]{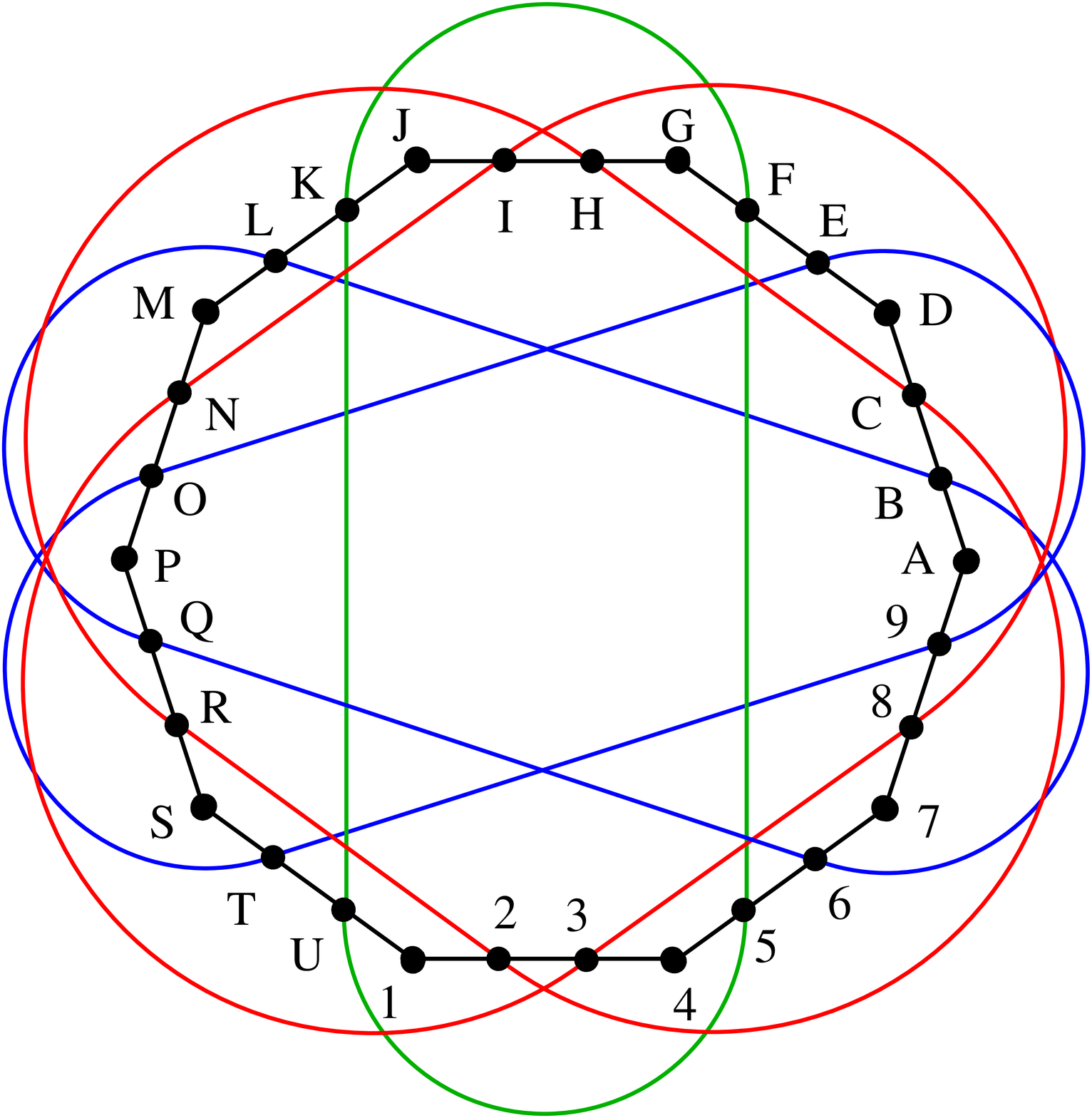}
\includegraphics[width=0.257\textwidth]{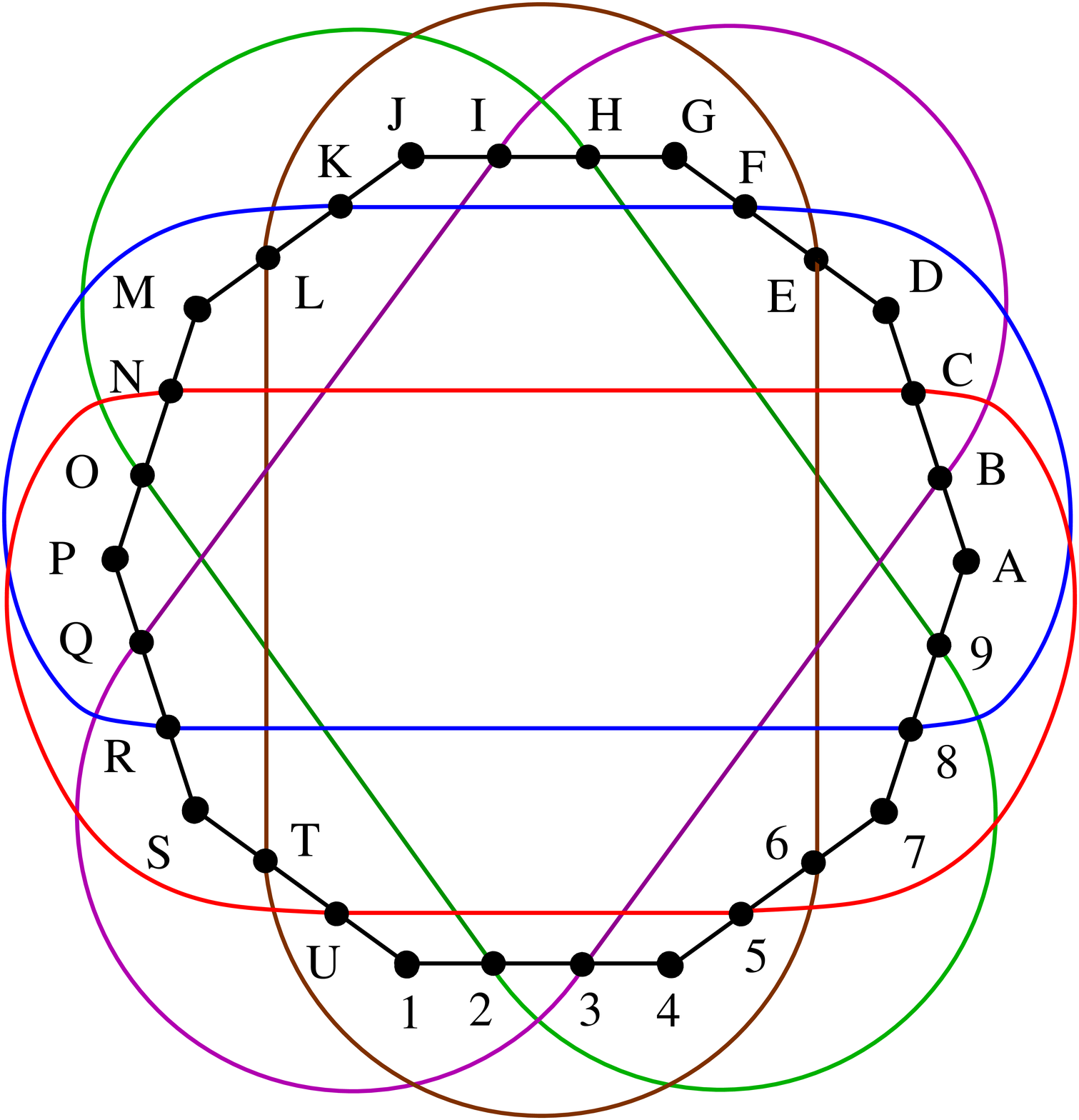}
\includegraphics[width=0.212\textwidth]{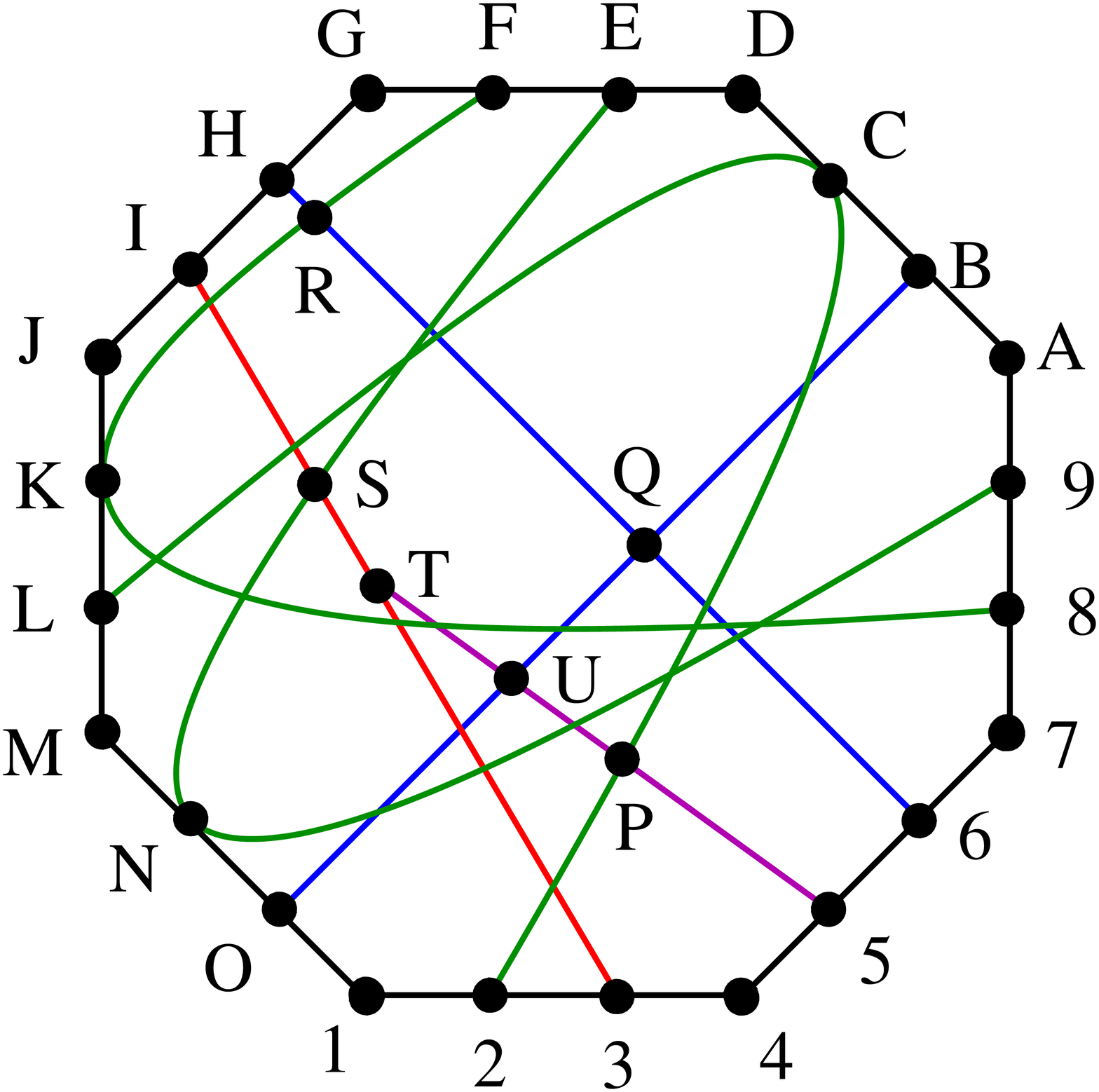}
\end{center}
\caption{[1st fig.] The smallest critical KS set, shown with the help of
MMP hypergraphs. It has 26 vertices, 13 edges, and a maximal
loop of 8 edges (octagon). Vertices correspond
to vectors and edges to orthogonal tetrads. [2nd-4th fig.]
Critical sets with 30 vertices, 15 edges, and
a maximal loop of 10 edges (decagon) for 2nd and 3rd and an octagon for the
4th figure. The  2nd one is isomorphic to the 1st and the 4th one to the 2nd
of the two 30-15 critical sets found in Ref.~\onlinecite{aravind10}, respectively }
\label{fig:c-26-30}
\end{figure}

Since the 24-24 KS set is a single largest set of its class, we expected
that sets and in particular critical ones of our new 60-75 class would be
based on loops larger than hexagons. The conjecture was correct.
Also, since by exhaustive generation of all
MMP hypergraphs up to 23 vertices we have not found a single
KS set with a loop larger than a hexagon, we concluded that
these classes of KS sets do not overlap, i.e., that the
minimal critical sets of the 60-75 KS sets must have more
than 24 vectors. See Fig.~\ref{fig:c-26-30}.

The figures in Fig.~\ref{fig:c-26-30} show MMP hypergraph representations
of the smallest critical subsets of the 60-75 KS set.  The smallest one has 26
vectors and 13 tetrads. Each vector is represented by a vertex in the MMP
hypergraph. Each tetrad, which consists of four mutually orthogonal vectors,
is represented by an edge in the MMP hypergraph. The next smallest ones
are three KS sets with  30 vectors and 15 tetrads. Their MMP hypergraphs
have 30 vertices and 15 edges each, as shown in their figures. As we can see,
the first three sets are highly symmetrical and might be suitable for setting up
an experiment.

In the MMP notation described above,
the 26-13 set has the following representation:

 {\tt 1234,4567,789A,ABCD,DEFG,GHIJ,JKLM,MNO1,5CHO,3Q8I,6QKF,NP9E,2PLB.}

As we can see, there is a one-to-one correspondence between this notation and
the figure. This is possible because in constructing a set,
we only deal with orthogonalities
between vectors and not with the values of the vector components. The values can always
be ascribed later on by means of our program {\tt vectorfind}.

The 60-75 \cite{aravind-600,aravind10} in the MMP notation reads:

{\noindent
{\tt  1234,1cKT,1Qtg,1Njo,1yYE,2Mmn,2vZD,2Pri,2bIV,3HWe,3kqO,3XGx,3shS,4Fwa,4UdJ,4fRu,4pLl,5678,5pSK,\break  5XiN,5buE,5Wwm,9ABC,9fxK,9sVN,9PlE,9qdZ,AUOt,AHiy,Abao,AGRm,BFSj,BXnc,BvJg,BWLr,CpeY,CkDQ,CMuT,\break ChwI,6Fet,6kVy,6PJo,6hLZ,7Uxj,7sDc,7Mag,7qRI,8fOY,8HnQ,8vlT,8Gdr,FGDE,FqiT,UhnE,UWVT,fhig,fWDo,\break pGVg,pqno,HIJK,HZuj,kraK,kmlj,XIlt,XZaY,srut,smJY,MLON,MdSy,vReN,vwxy,PRSQ,PwOc,bLxQ,bdec.}}

To obtain these results, we used an interactive procedure where we
stripped one block at a time and then decided (based e.g.\ on the size of
the output) what steps to take next.  The programs we used and their
algorithms are described in Sec.\ \ref{sec:alg}.  We used the program
{\tt mmpstrip} to strip blocks from starting diagrams.  In general, we
attempted to set the parameters of {\tt mmpstrip} (in particular, its
increment parameter) so that we ended up with a sample of about 10,000
hypergraphs after colorable hypergraphs and isomorphic hypergraphs
were removed.

The overall interactive procedure was as follows.  We start with
the MMP hypergraph for the 60-75 KS set.
\begin{enumerate}
\item  From this set, new sets with less and less blocks were 
  generated with with {\tt mmpstrip}.
  We used their increment parameter to keep the number of
  hypergraphs manageable, and we enabled the suppression of non-connected
  hypergraphs.
\item  Duplicate hypergraphs will result when one block is removed
  at a time (rather than multiple blocks combinatorially).
  These duplicates were removed.
\item
  Colorable hypergraphs were filtered out with {\tt states01},
  leaving only non-colorable ones (i.e.\ KS sets).
\item
  Isomorphic hypergraphs were removed with {\tt shortdL}.
\end{enumerate}
While it is not exhaustive, the advantage of the above technique
is that the filtering quickly converges to give us a collection of
non-isomorphic KS sets with the desired block count for further study.
Typically, we first ran these steps on  a small
sample of the hypergraphs (a hundred or so) so that the increment
parameter for {\tt mmpstrip} in the first step above could be estimated,
in order to end up with around 10,000 hypergraphs in the
last step.

In order to determine, after the above process (i.e.\ after removing a
single block then processing the output), which (non-colorable) hypergraphs
were critical, we ran {\tt mmpstrip} on one hypergraph at a time, then
determined with {\tt states01} whether all of them became colorable.  If
any one was non-colorable, it meant that the hypergraph was non-critical.
This procedure is somewhat CPU-intensive and for this study was done
(in conjunction with the above process)
only up to 19 blocks remaining (i.e.\ with 56 blocks or more blocks
removed).  We also did a more limited sampling of critical KS sets
with higher block counts.

\begin{figure}[htp]
\begin{center}
\includegraphics[width=0.24\textwidth]{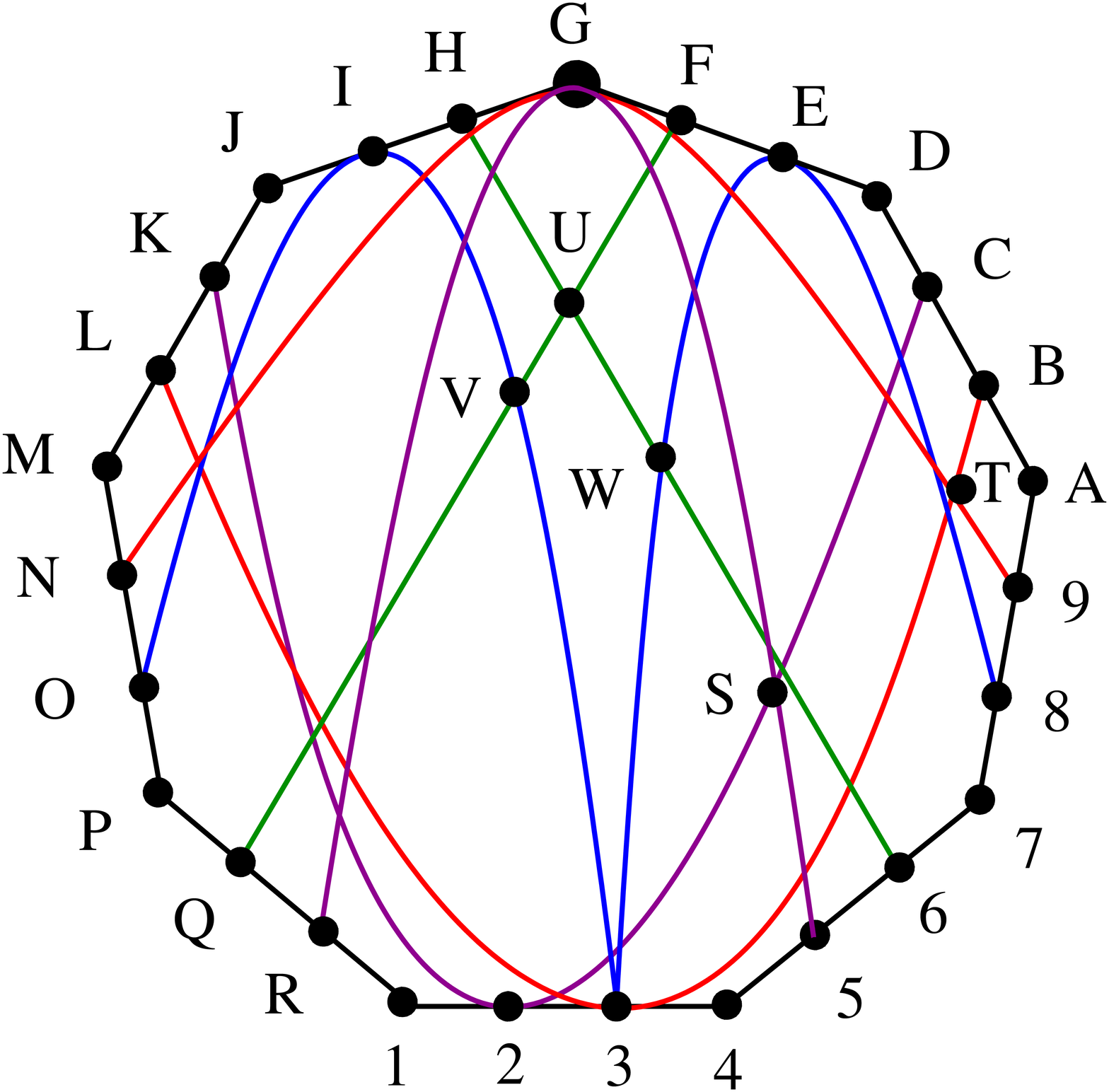}
\includegraphics[width=0.27\textwidth]{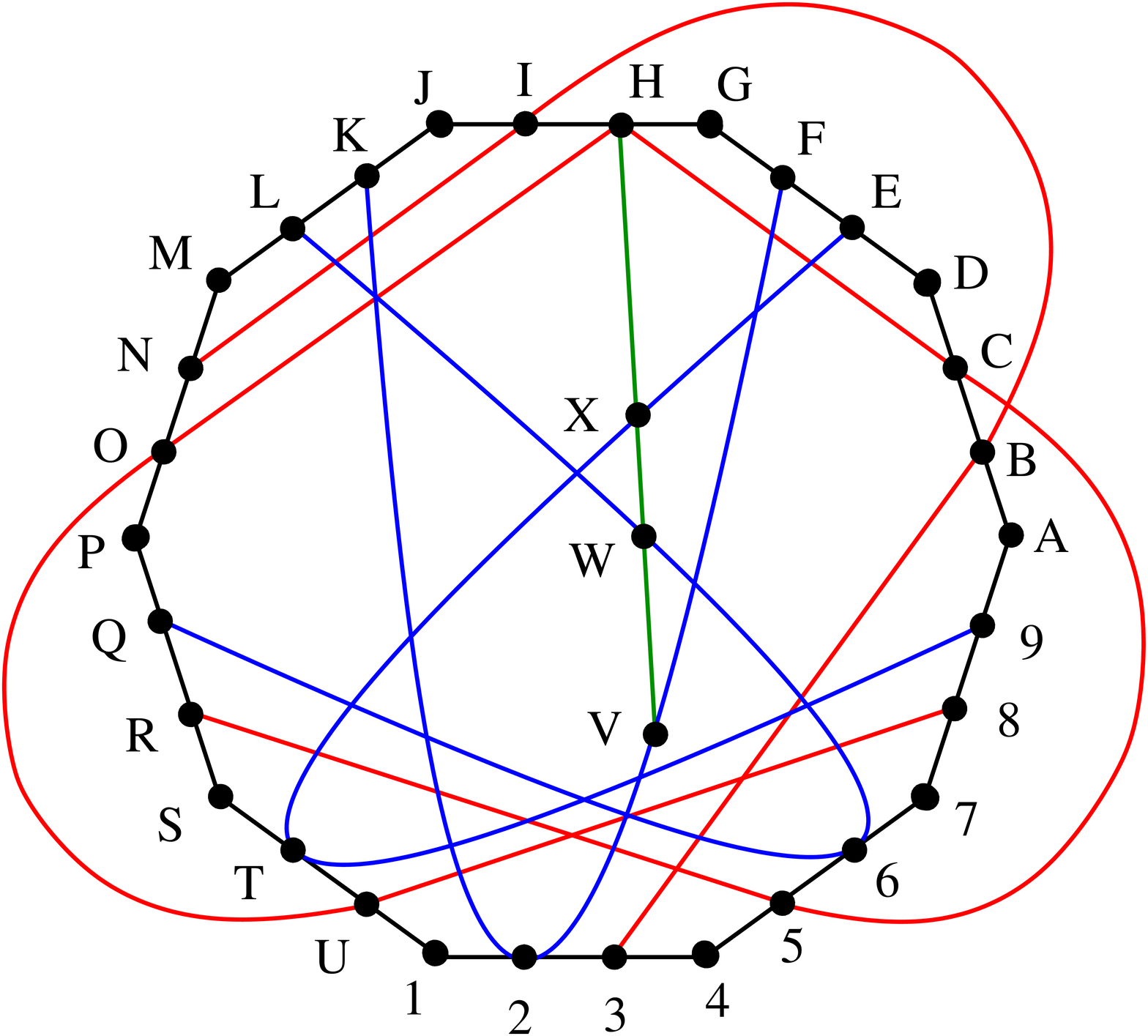}
\includegraphics[width=0.23\textwidth]{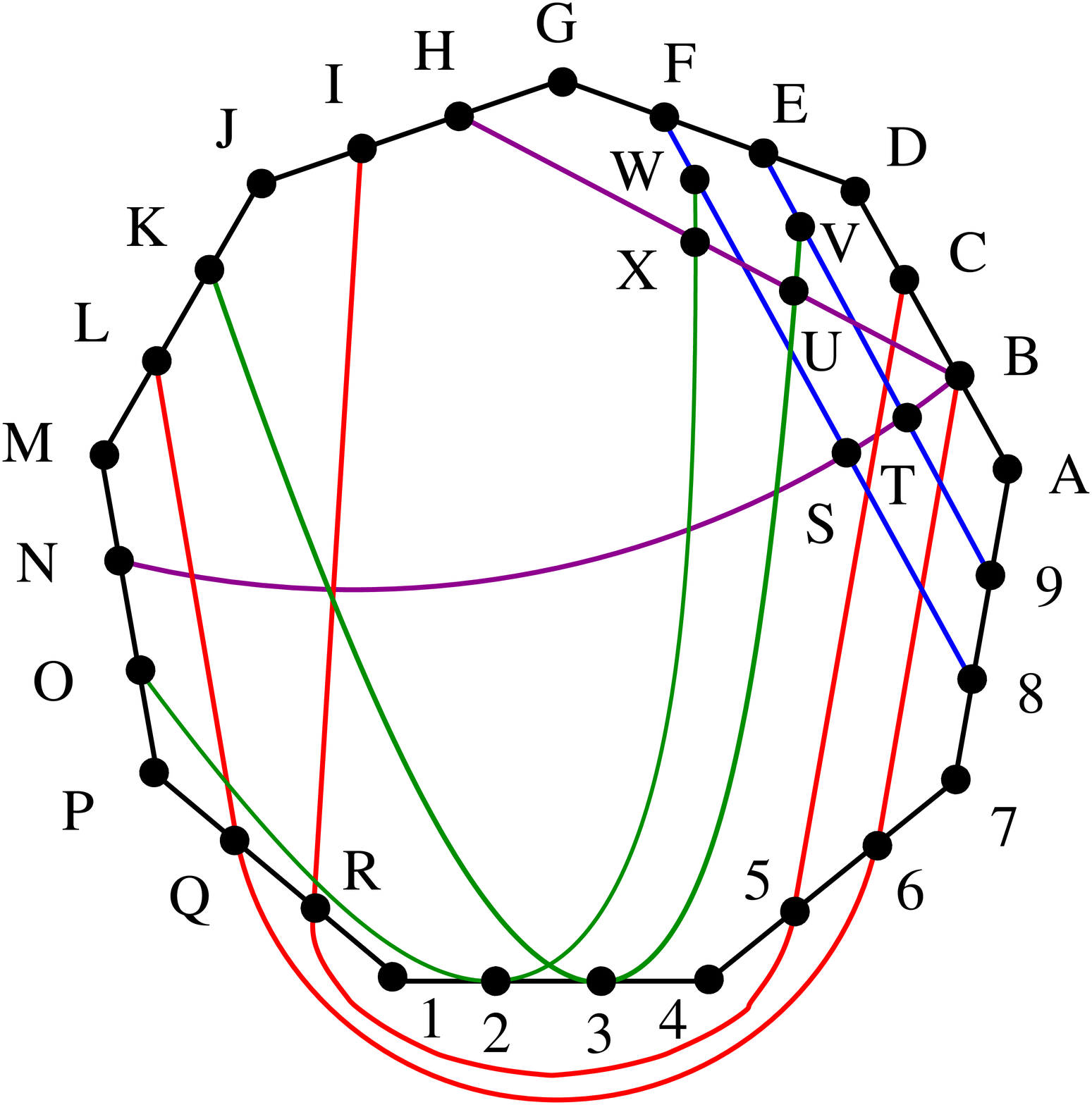}
\includegraphics[width=0.24\textwidth]{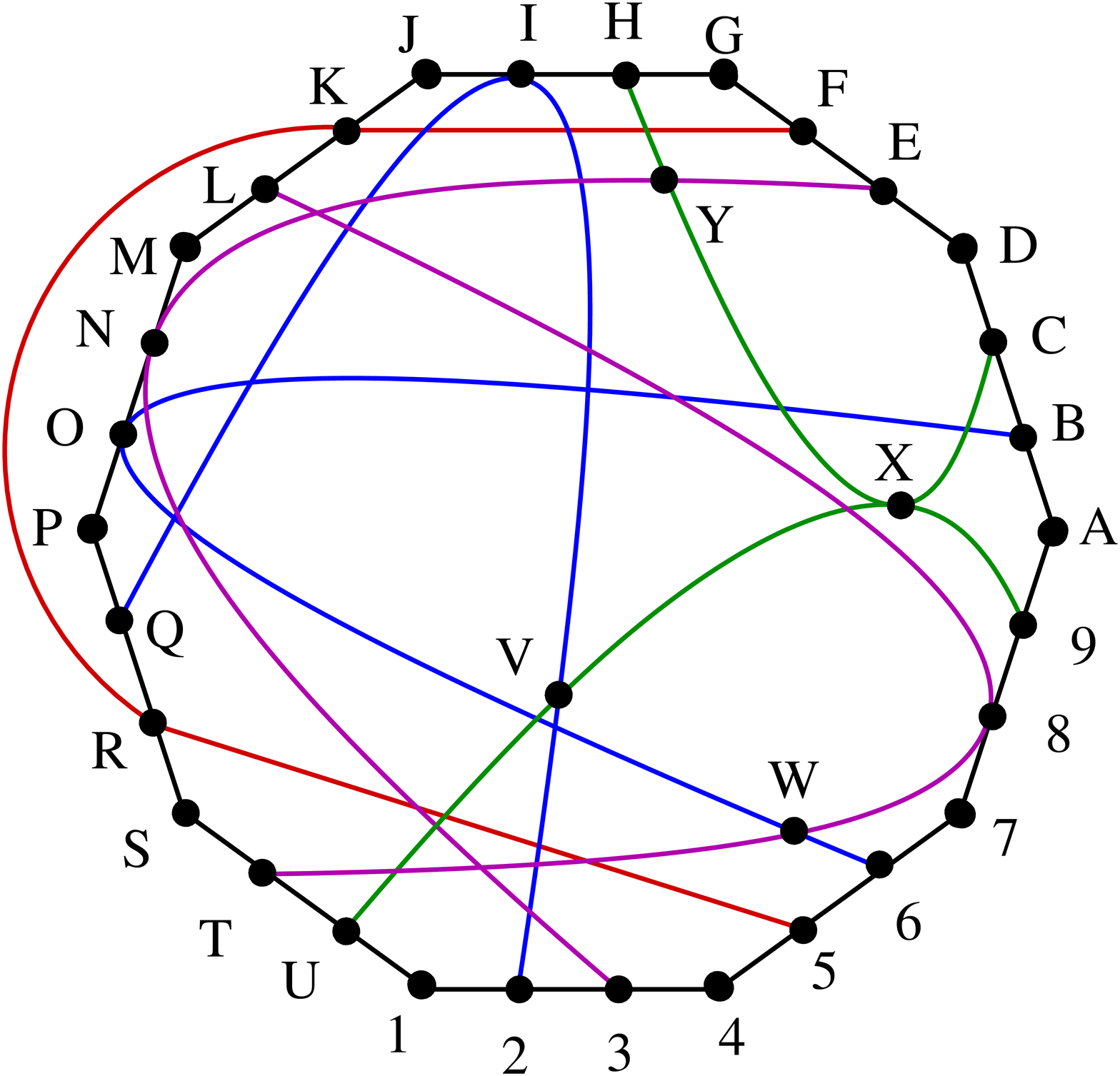}
\end{center}
\caption{[1st fig.] 32-17 KS set; Maximal loop: nonagon; [2nd, 3rd fig.] 33-17a,b; decagon, nonagon;
[4th fig.] 34-17a; decagon.}
\label{fig:c-32-34}
\end{figure}

The following list summarizes the critical hypergraphs we found with up
to 19 blocks, along with some sample MMP diagrams and figure references.
``30-15$\times$3'' means that we found 3 non-isomorphic critical
diagrams with 30 vectors and 15 blocks (tetrads).  This list resulted
from testing several million hypergraphs using the procedure described
above. The reader should keep in mind that the list is
not necessarily exhaustive but is based on this limited sample.


\begin{itemize}
\item $\le$ 12 blocks:  None.
\item 13 blocks:  26-13$\times$1.  It is shown in
Fig.\ \ref{fig:c-26-30}, with a maximal loop of 8 blocks (octagon).
Its MMP representation is:

{\noindent 26-13: \ \
{\tt 1234,4567,789A,ABCD,DEFG,GHIJ,JKLM,MNO1,5CHO,3Q8I,6QKF,NP9E,2PLB}}

\item 14 blocks:  None.
\item 15 blocks:  30-15$\times$3. The first two sets
in Fig.\ \ref{fig:c-26-30} hava a maximal loop of order 10 (decagon) and the
third one of order 8 (octagon). They receive the following MMP
representation:

{\noindent 30-15a: \ \
{\tt 1234,4567,789A,ABCD,DEFG,GHIJ,JKLM,MNOP,PQRS,STU1,5FKU,2CHR,38IN,9EOT,6BLQ.}}

{\noindent 30-15b: \ \
{\tt 1234,4567,789A,ABCD,DEFG,GHIJ,JKLM,MNOP,PQRS,STU1,6ELT,8FKR,C5UN,O29H,B3QI.}}

{\noindent 30-15c: \ \
{\tt 1234,4567,789A,ABCD,DEFG,GHIJ,JKLM,MNO1,2PCL,3TSI,5PUT,6QRH,8KRF,9NSE,BQUO.}}

30-15c is isomorphic to the first and 30-15a to the second of the two 30-15 critical sets found
in Ref.~\onlinecite{aravind10}.
\item 16 blocks:  None.
\item 17 blocks:  32-17$\times$1, 33-17 $\times$2, and 34-17 $\times$5.
The 32-17, 33-17b,  and 34-17d have maximal loops 9 (nonagon) and the 
maximal loopsof 33-17a-c and 34-17e are decagons, as shown in
Fig.~\ref{fig:c-32-34}.  Their MMP representations are

{\noindent 32-17: \ \
{\tt 1234,4567,789A,ABCD,DEFG,GHIJ,JKLM,MNOP,PQR1,2SCK,3VIO,3WE8,3LTB,5SGR,6WUH,9TGN,FUVQ.}}

{\noindent 33-17a: \ \
{\tt 1234,4567,789A,ABCD,DEFG,GHIJ,JKLM,MNOP,PQRS,STU1,K2VF,3BIN,R5CH,Q6WL,U8OH,9TXE,VWXH.}}

{\noindent 33-17b: \ \
{\tt 1234,4567,789A,ABCD,DEFG,GHIJ,JKLM,MNOP,PQR1,O2XW,K3UV,IR5C,LQ6B,8SWF,9TVE,BTSN,BUXH.}}

{\noindent 34-17a: \ \
{\tt 1234,4567,789A,ABCD,DEFG,GHIJ,JKLM,MNOP,PQRS,STU1,2VIQ,3NYE,5RKF,6WOB,TW8L,9XVU,CXYH.}}

{\noindent 34-17b: \ \
{\tt 1234,4567,789A,ABCD,DEFG,GHIJ,JKLM,MNOP,PQRS,STU1,2XHO,3CYR,5VKT,6BIN,8FWU,9ELQ,VWXY.}}

{\noindent 34-17c: \ \
{\tt 1234,4567,789A,ABCD,DEFG,GHIJ,JKLM,MNOP,PQRS,STU1,2WIN,3XHC,5FKU,6RVB,8YTO,9ELQ,VWXY.}}

{\noindent 34-17d: \ \
{\tt 1234,4567,789A,ABCD,DEFG,GHIJ,JKLM,MNOP,PQR1,O2T9,NS38,5UCH,6YBI,ELVQ,FKXR,XWUT,VSWY.}}

{\noindent 34-17e: \ \
{\tt 1234,4567,789A,ABCD,DEFG,GHIJ,JKLM,MNOP,PQRS,STU1,2VXI,3WYH,5FKU,6BOT,8VWR,9ELQ,CXYN.}}

\begin{figure}[htp]
\begin{center}
\includegraphics[width=0.23\textwidth]{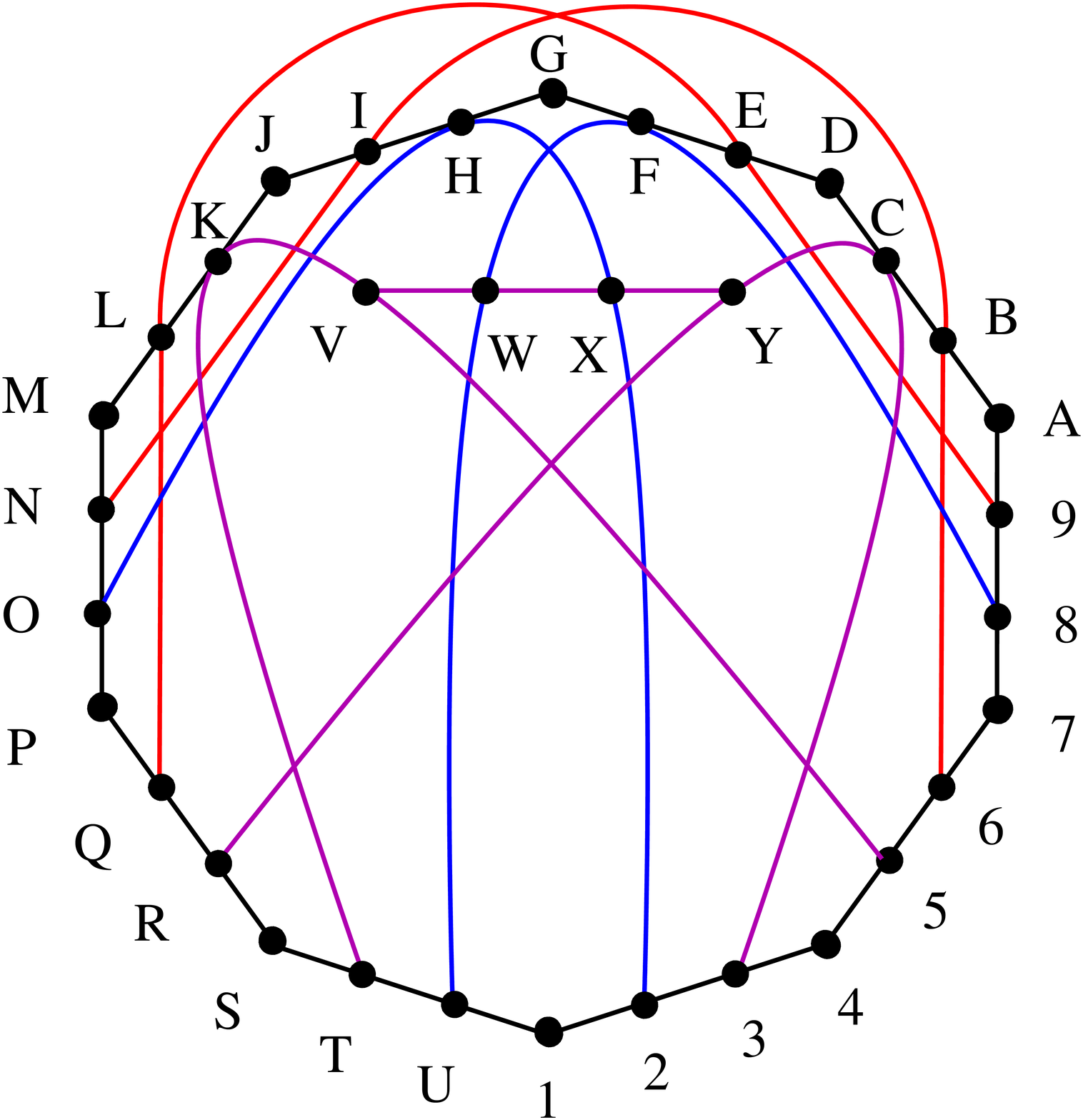}
\includegraphics[width=0.23\textwidth]{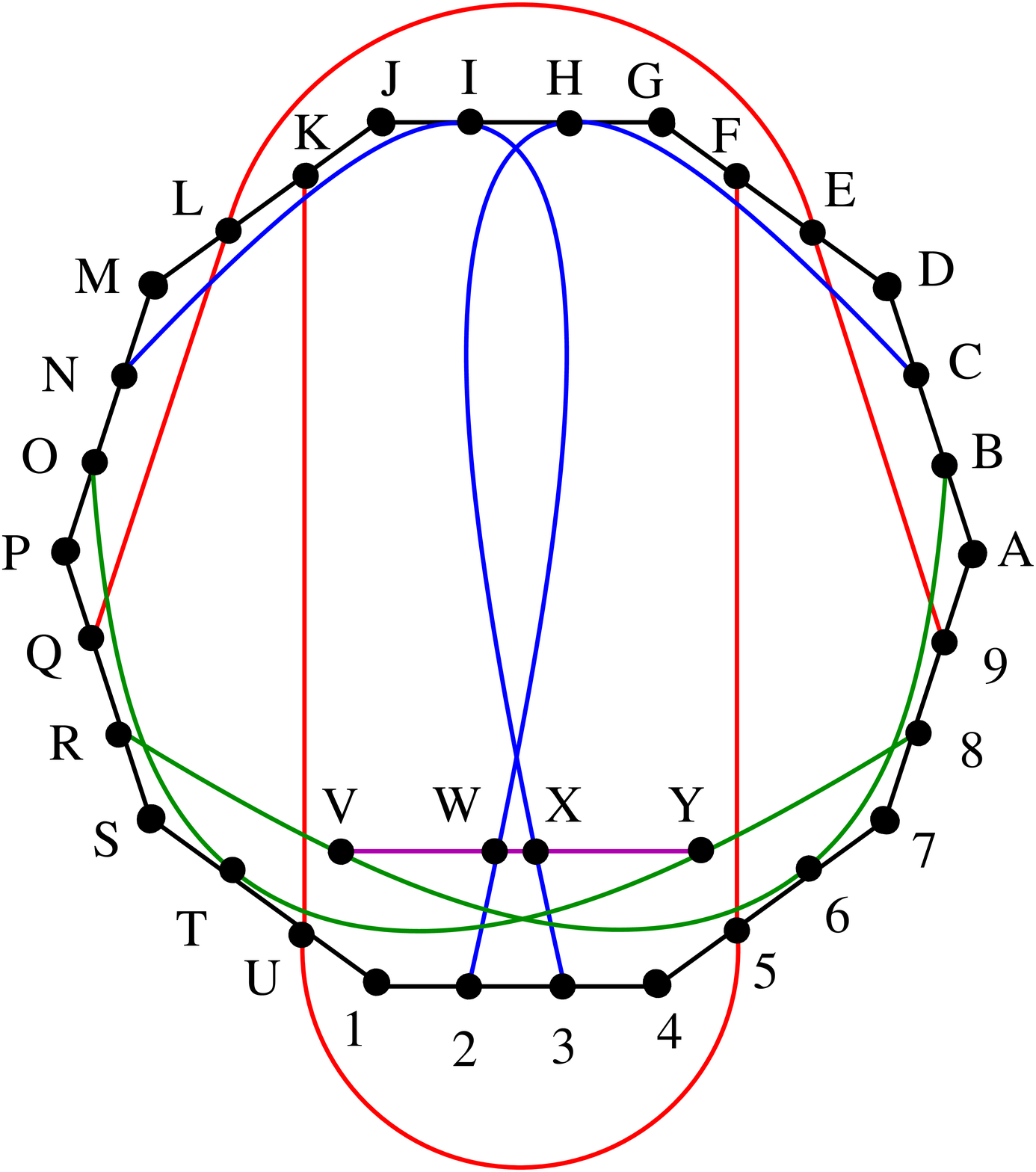}
\includegraphics[width=0.25\textwidth]{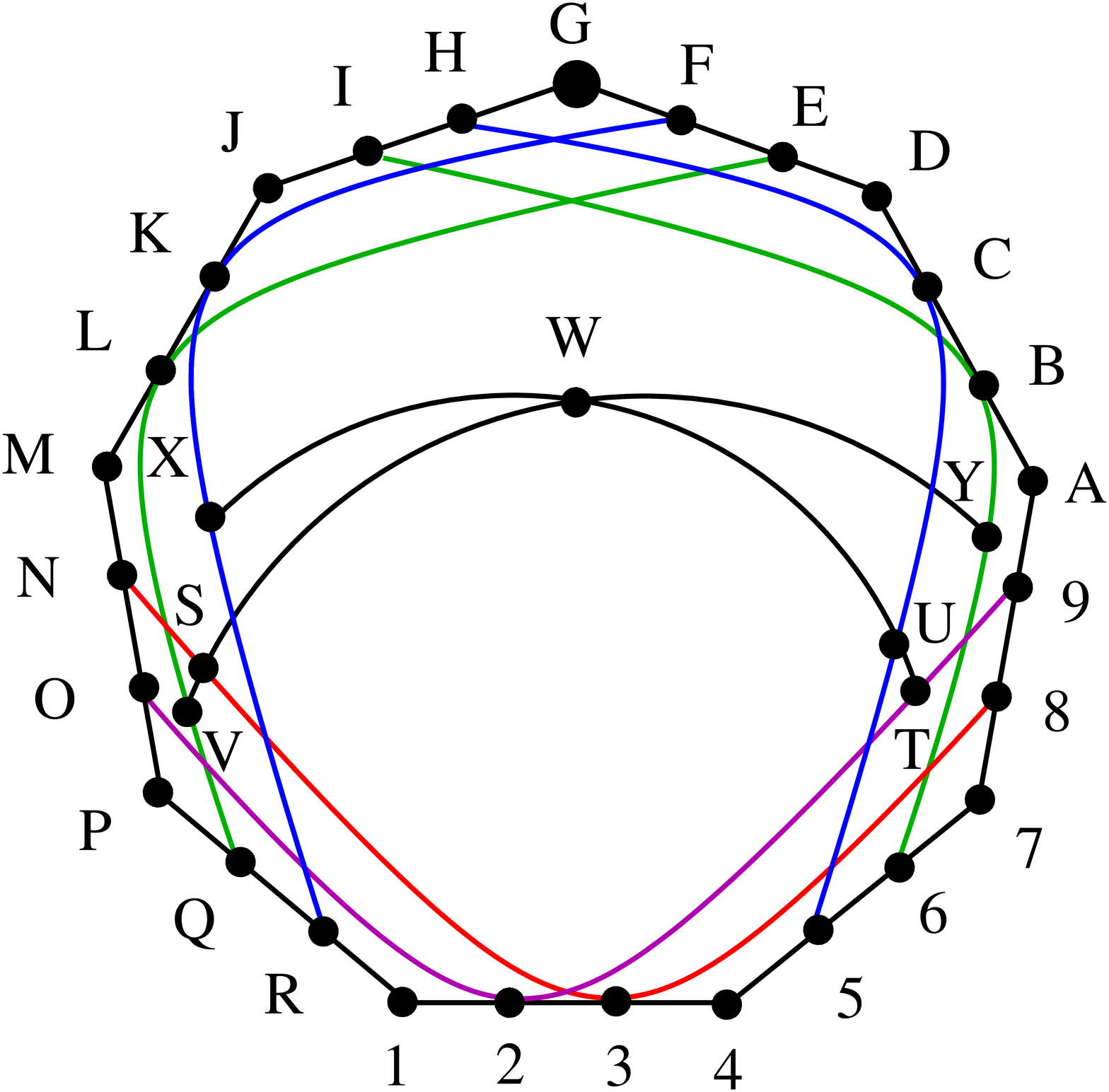}
\includegraphics[width=0.23\textwidth]{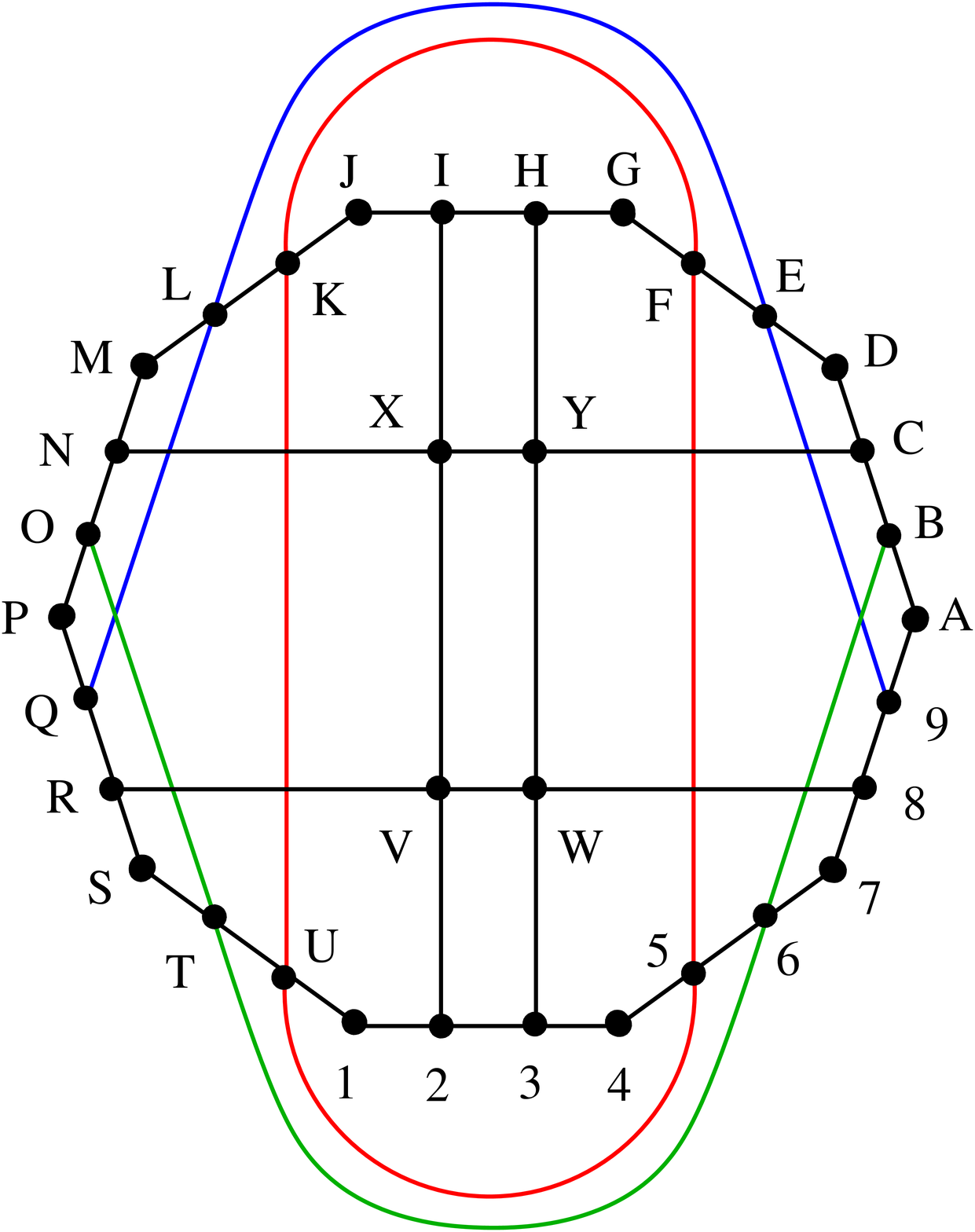}
\end{center}
\caption{[1st fig.] 34-17b (Max. loop: decagon); [2nd] 34-17c (decagon);
[3rd] 34-17d (nonagon); [4th] 34-17e (decagon).}
\label{fig:c-34-34}
\end{figure}

\item 18 blocks:  None.

\item 19 blocks:  36-19$\times$11, 37-19$\times$9, and 38-19$\times$6.
Three of them have the maximal loops of order 11.

\item 20 blocks: None

\item 21 and higher: The number of critical KS sets per block
increases to a maximum of over 1700 sets with 38 blocks, then decreases.
Even numbers of blocks in a KS set
start at 24. Odd numbers of vectors with an odd number of blocks
start to appear at 17 blocks
(33 vectors) and with an even number of blocks at 24 blocks.
The 33-17 set is given above, and
an MMP representation of a 42-24 set (with 14 blocks in a maximal loop) reads: \\
42-24:\phantom{xx}{\tt 1234,5678,9ABC,DEF8,GHIJ,KLMJ,NOPQ,RSTM,UVI4,WXYF,ZaE3,bcLC,dTHB,eaYQ,ecVD,fbPG,gUSO,\\
\phantom{xxxXxXx}gfeA,ZTN7,XU97,bWR2,dWO6,fK63,eJ72.}\\
An exhaustive generation proves that there are no critical sets with more than
63 blocks. We conjecture that the first critical sets will appear at
60 blocks, i.e., for 60-60 KS sets and smaller,
in analogy to the 24-24 class, assuming that
that picking up new tetrads (61st-63rd) of the already recorded
 60 vector/state values
would not make a set empirically distinguishable from the 60-60
one we started with.
The size of maximal loops steadily rises with the number of blocks.
(Detailed presentation of additional KS sets we obtained
by two other techniques will be given
elsewhere, because the algorithms
we use for higher numbers of blocks are too involved to be presented here.
Besides, many massive
computations that take months on our clusters are under way.)
\end{itemize}

Since our sampling was not exhaustive---for example, there are 290 quadrillion
(2.9$\times 10^{17}$) hypergraphs with 19 blocks---it is likely that
there are many more critical KS sets than suggested above, particularly
at block counts higher than 19.  When
repeating the above procedure with independent random samples, some
critical diagrams (up to isomorphism) recurred frequently, whereas
others occurred only once or twice, indicating that their distribution
and symmetries are far from uniform.  An exhaustive generation of all sets from
the 60-75 KS class, that we might be able to carry out in the future, would
give us the complete list.

\section{\label{sec:alg}Algorithms}

Our study makes use of algorithms and computer programs from several
disciplines (quantum mechanics, lattice theory, graph theory, and
geometry).  Each has its own terminology, which we will sometimes keep
when discussing an algorithm from that discipline.  To avoid confusion, the
reader should keep in mind that in the context of the MMP diagrams used
for this study, the terms ``vertex,'' ``atom,'' ``ray,'' ``1-dim
subspace,'' and ``vector'' are synonymous, as are the terms ``edge,''
``block,'' and ``tetrad (of mutually orthogonal vectors).''
Similarly, ``MMP hypergraph'' and ``MMP diagram'' mean the same
thing.

For the purpose of the KS theorem, the vertices of an MMP hypergraph are
interpreted as rays, i.e.\ 1-dim subspaces of a Hilbert space, each
specified by a representative (non-zero) vector in the subspace.  The
vertices on a single edge are assumed to be mutually orthogonal rays or
vectors.  In order for an MMP hypergraph to correspond to a KS set,
first there must exist an assignment of vectors to the vertices such
that the orthogonality conditions specified by the edges are satisfied.
Second, there must not exist an assignment (sometimes called a
``coloring'') of 0/1 (non-dispersive or classical) probability states to
the vertices such that each edge has exactly one vertex assigned to 1
and its others assigned to 0.

For a given MMP hypergraph, we use two programs to confirm these two
conditions.  The first one, {\tt vectorfind}, attempts to find an
assignment of vectors to the vertices that meets the above requirement.
This program is described in Ref.~\onlinecite{pmm-2-09}.  The second program,
{\tt states01}, determines whether or not a 0/1 coloring is possible
that meets the above requirement.  The algorithm used by {\tt states01}
is described in Ref.~\onlinecite{pmmm03a}.

The 60-vertex, 75-edge MMP hypergraph based on the 600-cell described above
(which we refer to as 60-75) has been shown to be a KS set.\
\cite{aravind-600} However, we can
remove blocks from it and it will continue to be a KS set.  The
purpose of this study was to try to find subsets of the 60-75 hypergraph
that are critical i.e.\ that are minimal in the sense that if any one
block is removed, the subset is no longer a KS set.

While the program {\tt vectorfind} independently confirmed that 60-75
admits the necessary vector assignment, such an assignment remains valid
when a block is removed.  Thus it is not necessary to run {\tt
vectorfind} on subsets of 60-75.  However, a non-colorable (KS) set
will eventually admit a coloring when enough blocks are removed, and the
program {\tt states01} is used to test for this condition.

The basic method in our study was to start with the 60-75 hypergraph and
generate successive subsets, each with one or more blocks stripped off
of the previous subset, then keep the ones that continued to admit no
coloring and discard the rest.  Of these, ones isomorphic to others were
also discarded.

The program {\tt mmpstrip} was used to generate subsets with blocks
stripped off.  The user provides the number of blocks $k$ to strip from
an input MMP hypergraph with $n$ blocks, and the program will produce all
${{n}\choose{k}}$ subsets with a simple combinatorial algorithm.
Partial output sets can be generated with start and end parameters, and
if the full output is too large to be practical, an increment parameter
$i$ will skip all but every $i$th output line in order to partially
sample the output subsets.  Given an input file with MMP hypergraphs, the
program can calculate in advance how many output hypergraphs will result,
so that the user can plan which parameter settings to use.

The {\tt mmpstrip} program will also optionally suppress MMP hypergraphs
that are not connected, such as those with isolated blocks or two
unconnected sections, since these are of no interest.  Finally, all
output lines are by default renormalized (assigned a canonical atom
naming), so that there are no gaps in the atom naming as is required by
some other MMP processing programs.

In order to detect isomorphic hypergraphs, one of
two programs was used.  For testing small sets of
hypergraphs, we used the program {\tt subgraph}
described in Ref.~\onlinecite{pmm-2-09}, which has the advantage of
displaying the isomorphism mapping for manual verification.
For a large number of hypergraphs, we used
Brendan McKay's program {\tt shortdL}, which has a much faster
run time.

\section{\label{sec:disc}Discussion}

In this paper we describe the 60-75 class of Kochen-Specker
set with a focus on its smallest critical sets, as defined in the
Introduction. The smallest critical sets we find are shown
in  Figs.~\ref{fig:c-26-30}-\ref{fig:c-34-34}. The order of their maximal loops of
edges (blocks, tetrads) is 8 and more. Since the order of the maximal
loop of all subsets that form the lower KS 24-24 class is 6, this is an
additional aspect of the disjointness of 24-24 and 60-75 classes, for which
we have shown to have a statistical confidence of over 95\%.
More details on the latter results are given in the Introduction.

The high symmetry of the smallest critical KS sets shown in
the first three figures of Fig.~\ref{fig:c-26-30} and in the figures of
Fig.~\ref{fig:c-34-34} suggests that spin KS experiments might be
designed for them. Therefore we would like to discuss geometrical features
of the sets we obtained in Sec.~\ref{sec:critical}.

Each of the sets shown in  Figs.~\ref{fig:c-26-30}-\ref{fig:c-34-34}
involves an odd number of bases (blocks, edges, tetrads)  (13, 17 or 17),
with each ray (vertex, vector, direction) occurring exactly twice over
these bases.  This observation, by itself, gives an immediate ``parity
proof'' of the BKS (Bell-Kochen-Specker) theorem without the need
for any further calculation or analysis (and, in particular, without the
need to use program {\tt states01}  mentioned in the previous section).
The reason is the following:  on the one hand, because each
tetrad must have exactly one ray assigned the value 1, there must be an
odd number of occurrences of 1's over all the tetrads; but, on the other
hand, because each ray is repeated twice, there must be an even number
of 1's over all the tetrads.  This contradiction shows that a 1/0
assignment is impossible and so proves that these are indeed KS sets.
The argument, of course, does not go through for those sets where
a ray appears in an odd number of tetrads as, e.g., ray 2 in the the set
42-24 whose MMP representation is given at the end of
Sec.~\ref{sec:critical}.

An interesting difference between the 26-13 and 30-15 cases is that
the latter are isogonal (or vertex transitive), whereas the former is
not.  A set of rays is said to be isogonal if there is a symmetry
operation that will take any ray into any other one while keeping the
structure as a whole invariant.  The 60 rays of the 600-cell are
isogonal as a whole, and this might encourage the belief that subsets
yielding parity proofs must also be isogonal.  The 26-13 set shows this
supposition to be false in the case of the 600-cell, or the 60-75 set.

In addition to the methods outlined in the previous sections,
alternative methods were used to arrive at some of the possible critical
sets. The idea, which followed the ``parity proof'' above, was to look
for $N$ rays forming $T$ complete tetrads, with  $T$ odd, in such a
way that each ray occurred in exactly two of the tetrads.
Such a set, which we will refer to as a $N$-$T$ set
(e.g., 26-13, 30-15, etc.), is a KS set.  An $N$-$T$ set that obeys the
``parity proof'' satisfies the numerical constraint $N=2T$, so it
involves only one free parameter.  (Of course, there are other
critical sets such as 33-17 that will not be found by this method.)
Starting from small values of $N$ and proceeding upwards, we looked
for KS sets.  It is easily seen that no set of this type can exist for $N$
less than 16.

The reason has to do with the structure of the tetrads for the 600-cell.
Inspection shows that each ray occurs in exactly five tetrads and that it
occurs exactly once in these tetrads with
each of the 15 rays it is orthogonal to.
Suppose a particular tetrad is chosen as
the ``seed'' for a $N$-$T$ set.  Then
each ray in that tetrad must occur in one other tetrad, and so there
must be at least four other tetrads involved.  However each of those
tetrads must involve three new rays, and so the total number of rays,
including the four we began with, is 16.  Starting with $N = 18$ and
proceeding upwards (remembering that $T = N/2$ must be an odd integer)
shows, through slightly more involved arguments (which differ from those
for the smallest critical set (18-9) of the 24-24 class),
that solutions with $N = 18$ and 22 are impossible.  The first solution
that is possible is for $N = 26$, and it explains the 26-13 set shown in
Fig.\ \ref{fig:c-26-30}.  There are actually 1800 different sets of 26 rays
that lead to such a solution, but they are all geometrically isomorphic
to one another, in the sense that there is a four-dimensional orthogonal
transformation that will take any one such set into any other.  In the
Introduction, we give a statistical argument showing, with over 95\%\
confidence, that there are no smaller sets than
26-13 in the 60-75 class that do not follow the parity proof.

All the other results we obtained, together with the presentation of the
algorithms and programs we used, is given in
Ref.~\onlinecite{waeg-aravind-megill-pavicic-10}. There we give a
detailed analyisis of the results, complete statistics of the obtained
sets, and a review of their features. All that is outside of the scope of
the present paper.

\begin{acknowledgments}
One of us (M. P.) would like to thank his host Hossein Sadeghpour
for a support during his stay at ITAMP.

Supported by the US National Science Foundation through a
grant for the Institute for Theoretical Atomic, Molecular,
and Optical Physics (ITAMP) at Harvard University and Smithsonian
Astrophysical Observatory and Ministry of
Science, Education, and Sport of Croatia through the project
No.\ 082-0982562-3160.

Computational support was provided by the cluster Isabella of
the University Computing Centre of the University of Zagreb and
by the Croatian National Grid Infrastructure.
\end{acknowledgments}

\end{document}